\documentclass[a4paper,10pt]{article}
\usepackage{graphicx,fullpage,cite}
\usepackage{amsfonts,amsmath,amssymb}
\usepackage[utf8x]{inputenc}
\title{Quantum Jarzynski Equality with multiple measurement and feedback for isolated system}
\author{Shubhashis Rana, Sourabh Lahiri, A. M. Jayannavar}
\date{}

\allowdisplaybreaks
\begin{document}

\newcommand{\nwc}{\newcommand}
\nwc{\la}{\langle}
\nwc{\ra}{\rangle}
\nwc{\lw}{\linewidth}
\nwc{\nn}{\nonumber}
\nwc{\Ra}{\Rightarrow}
\nwc{\dg}{\dagger}

\nwc{\Tr}[1]{\underset{#1}{\mbox{\large Tr}}~}
\nwc{\pd}[2]{\frac{\partial #1}{\partial #2}}
\nwc{\ppd}[2]{\frac{\partial^2 #1}{\partial #2^2}}

\nwc{\zprl}[3]{Phys. Rev. Lett. ~{\bf #1},~#2~(#3)}
\nwc{\zpre}[3]{Phys. Rev. E ~{\bf #1},~#2~(#3)}
\nwc{\zpra}[3]{Phys. Rev. A ~{\bf #1},~#2~(#3)}
\nwc{\zjsm}[3]{J. Stat. Mech. ~{\bf #1},~#2~(#3)}
\nwc{\zepjb}[3]{Eur. Phys. J. B ~{\bf #1},~#2~(#3)}
\nwc{\zrmp}[3]{Rev. Mod. Phys. ~{\bf #1},~#2~(#3)}
\nwc{\zepl}[3]{Europhys. Lett. ~{\bf #1},~#2~(#3)}
\nwc{\zjsp}[3]{J. Stat. Phys. ~{\bf #1},~#2~(#3)}
\nwc{\zptps}[3]{Prog. Theor. Phys. Suppl. ~{\bf #1},~#2~(#3)}
\nwc{\zpt}[3]{Physics Today ~{\bf #1},~#2~(#3)}
\nwc{\zap}[3]{Adv. Phys. ~{\bf #1},~#2~(#3)}
\nwc{\zjpcm}[3]{J. Phys. Condens. Matter ~{\bf #1},~#2~(#3)}
\nwc{\zjpa}[3]{J. Phys. A ~{\bf #1},~#2~(#3)}
\nwc{\zpjp}[3]{Pram. J. Phys. ~{\bf #1},~#2~(#3)}
\nwc{\zpa}[3]{Physica A ~{\bf #1},~#2~(#3)}

\maketitle{}
  \begin{center}
    Institute of Physics, Bhubaneswar - 751005, Sachivalaya Marg, India.
  \end{center}
  \begin{abstract}

    In this paper, we derive the Jarzynski equality (JE) for an isolated quantum system in three different cases: (i) the full evolution is unitary with no intermediate measurements, (ii) with intermediate measurements of arbitrary observables being performed, and (iii) with intermediate measurements whose outcomes are used to modify the external protocol (feedback). We assume that the measurements will involve errors that are purely  classical in nature.  Our treatment is based on  path probability in state space for each realization. This is in  contrast to the formal approach based on projection operator and density matrices. We find that the JE remains unaffected in the second case, but gets modified in the third case where the mutual information between the measured values with the actual eigenvalues must be incorporated into the relation.

\vspace{0.5cm}
  \end{abstract}

\vspace{1cm}

\section{Introduction}

In the last couple of decades a lot of work has been directed towards nonequilibrium statistical mechanics, and has given birth to several equalities that are valid even when the system is far from equilibrium. They are collectively known as the \emph{fluctuation theorems} \cite{eva93,eva94,jar97,cro98,cro99}. These theorems also shed new light on some fundamental problems such as how irreversibility arises from underlying time-reversible dynamics. Moreover, these theorems will have important application in nanotechnology and nano physics. One of the pioneering works was due to Jarzynski \cite{jar97}, who had derived a relation between the nonquilibrium work performed on a system to change in its equilibrium free energy. Let us consider a system that is initially at canonical equilibrium with a heat bath at inverse temperature $\beta=\frac{1}{k_{B}T}$. Subsequently an external perturbation $\lambda(t)$, called  protocol, is applied to the system that takes it out of equilibrium. At time $t=\tau$, the process is terminated when the parameter value reaches $\lambda(\tau)$. The work $W$ done on the system will in general vary for different phase space trajectories, owing to the randomness of the initial state and thermal fluctuations due to coupling with the environment during the evolution. The Jarzynski equality (JE) states that,
\begin{equation}
 \langle e^{-\beta W}\rangle = e^{-\beta \Delta F}.
\end{equation}
Here, the angular brackets denote ensemble averaging over a large number of repetitions of the experiment. $\Delta F \equiv F(\lambda(\tau))-F(\lambda(0))$ is the difference in the equilibrium free energy of the system between the final and the initial states.
The JE has been extended to quantum domain \cite{han11b} in presence of measurement \cite{han10} and feedback \cite{sag08,tas11}. JE in presence of feedback has also been verified experimentally \cite{sag10a}. Quantum feedbacks are important in nanosytems or mesoscopic systems and can be applied to produce the cooling of nanomechanical resonators and atoms \cite{jac03,jac04}.

In our present study we derive quantum extended JE with multiple measurements and feedback for an isolated system. Our treatment is  based on path probability in state space for each realization as opposed to formal approach dealing with projection operator and density matrices \cite{sag08,tas11}. All the results are simple extensions of the theorems for fixed protocol, and the latter in turn depends on the principle of microscopic reversibility. 

For the quantum case to obtain the work values, we perform measurement (von Neumann type) of system energies (or Hamiltonian $H(t)$) at the beginning and end of protocol. The measured energy eigenvalues are denoted by $E_{i_{0}}(\lambda(0))$ and $E_{i_{\tau}}(\lambda(\tau))$ and corresponding instantaneous eigenstates by $|i_{0}\ra$ and $|i_{\tau}\ra$ respectively. The work done on the system by changing external protocol $\lambda(t)$ is given by
\begin{equation}
 W=E_{i_{\tau}}(\lambda(\tau))-E_{i_{0}}(\lambda(0)).
\label{work}
\end{equation}
$W$ is a realization dependent random variable. Initially the system is brought into contact with large reservoir at temperature T, thereby allowing the system to equilibrate. Subsequently the system is decoupled from the bath and the system evolves unitarily with a given Hamiltonian H(t). Our treatment closely follows \cite{sag11} wherein Hamiltonian derivation of JE
under feedback control is derived for classical case.

 Probability of system being in state $|i_{0}\ra$ is given by 
\begin{equation}
p(i_{0})=\dfrac{e^{-\beta E_{i_{0}}(\lambda (0)) }}{Z_{0}}.
\label{iprob}
\end{equation}
 The partition function is defined as
\begin{equation}
 Z_0 = \sum_{i_{0}} e^{-\beta E_{i_{0}}(\lambda (0))}.
\end{equation}
Between measurements, the system undergoes unitary evolution with an operator U given by
\begin{equation}
 U_{\lambda}(t_{2},t_{1})=T \exp\left( -\frac{i}{\hbar}\int^{t_{2}}_{t_{1}}H(t,\lambda(t))dt\right),
\end{equation}
where T denotes time ordering and $H(t)$ is the system Hamiltonian. The probability of the system initially in the state $|i_{0}\ra$ to be found in state $|i_{\tau}\ra$ at time $\tau$ is given by 
\begin{equation}
 P(i_{\tau}|i_{0})=|\langle i_{\tau}|U_{\lambda}(\tau,0)|i_{0}\rangle|^{2} .
\label{cprob1}
\end{equation}
Thus the joint probability of state being in $|i_{0}\ra$ and $|i_{\tau}\ra$ is
\begin{equation}
 P(i_{\tau},i_{0})=P(i_{\tau}|i_{0})p(i_{0})~~~~~~~\mbox{(Bayes' theorem)}
\label{jprob1}
\end{equation}
In  section \ref{sec:JE}, we rederive the JE for a quantum particle to make the paper self consistent. In section \ref{sec:JEm}, we derive the same with measurements of arbitrary observables being performed in-between. In section \ref{sec:JEf}, we derive the extended JE for a system with the protocol being monitored by a feedback control that changes the protocol according to the outcomes of the measurements performed. In section \ref{sec:effpara} generalized JE involving efficacy parameter is derived.

\section{Jarzynski Equality}\label{sec:JE}

For deriving JE we need to calculate  $\langle e^{-\beta W}\rangle$ which is given by
\begin{equation}
\la e^{-\beta W}\ra =\sum _{i_{\tau},i_{0}}e^{-\beta W} P(i_{\tau},i_{0}).
\end{equation}
Substituting the expression for realization dependent work (Eq.(\ref{work})) and joint probability $P(i_{\tau},i_{0})$ (Eq.(\ref{jprob1})) and using Eq.(\ref{iprob}) and Eq.(\ref{cprob1}) we get
\begin{align}
\langle e^{-\beta W}\rangle
&=\sum_{i_{0},i_{\tau}}e^{-\beta(E_{i_{\tau}}(\lambda(\tau))-E_{i_{0}}(\lambda(0))}  |\langle i_{\tau}|U_{\lambda}(\tau,0)|i_{0}\rangle|^{2}\dfrac{e^{-\beta E_{i_{0}}(\lambda (0)) }}{Z_{0}} \nn\\
&=\sum_{i_{0},i_{\tau}}\dfrac{e^{-\beta E_{i_{\tau}}(\lambda (\tau)) }}{Z_{0}} 
\langle i_{\tau}|U_{\lambda}(\tau,0)|i_{0}\rangle\langle i_{0}|U^{\dagger}_{\lambda}(\tau,0)|i_{\tau}\rangle\nn\\
\end{align}
Making use of completeness relation $\sum_{i_{0}}|i_{0}\ra\la i_{0}|=1$ and normalization condition $\la i_{\tau}|i_{\tau}\ra =1$ and unitarity of evolution, $U_{\lambda}^{\dagger}U_{\lambda}=1$, we have,
\begin{align}
\langle e^{-\beta W}\rangle 
=\sum_{i_{\tau}}\dfrac{e^{-\beta E_{i_{\tau}}(\lambda (\tau)) }}{Z_{0}} 
=\dfrac{Z_{\tau}}{Z_{0}}
=e^{-\beta \Delta F}.
\end{align}
where,  $Z_{\tau}=\sum_{i_{\tau}}e^{-\beta E_{i_{\tau}}(\lambda (\tau))}$, is the partition function of the system with the control parameter held fixed at $\lambda(\tau)$ and $\Delta F=\ln\frac{Z_{0}}{Z_{\tau}}$ is the equilibrium free energy difference between final and initial \index{}states. This is the quantum version of the JE \cite{han11b}. Using Jensen's inequality, we retrieve the second law from the above relation:
\begin{equation}
\la W\ra \ge \Delta F,
\end{equation}
implying second law is valid for average $W$ although for some individual realizations, $W$ can be less than $\Delta F$.
\section{JE in presence of measurement}\label{sec:JEm}
 This time, one intermediate measurement (of arbitrary observables, not necessarily the Hamiltonian) at time $t_{1}$ has been carried out but the entire protocol $\lambda(t)$ is predetermined. At time $t_{1}$ the state collapses to $|i_{1}\rangle$ after which it evolves according to the  unitary operator $U_{\lambda}(\tau,t_{1})$ up to the final time $\tau$. It is to be noted that the projective measurements  result in collapse of the system state to one of the eigenstates. This leads to decoherence and dephasing in further quantum evolution. If along two paths, intermediate measurements are performed, then the interference between alternative paths disappear and quantum effects are suppressed. Hence in presence of measurement, path probabilities in state space obeys simple classical probability rules. For example,  the path probability  is simply the product of the  transition probabilities between subsequent measured states. However, it may be noted that quantum mechanics enters through the explicit calculation of transition probabilities between states.
 The joint probability of the state trajectory is
\begin{align}
 P(i_{\tau},i_{1},i_{0})
&=p(i_{\tau}|i_{1})p(i_{1}|i_{0})p(i_{0})\\
&=|\langle i_{\tau}|U_{\lambda_{y_{1}}}(\tau,t_{1})|i_{1}\rangle|^{2} 
|\langle i_{1}|U_{\lambda}(t_{1},0)|i_{0}\rangle|^{2}p(i_{0}).
\label{jprob2}
\end{align}
Then,
\begin{equation}
 \langle e^{-\beta W}\rangle 
=\sum _{i_{\tau},i_{1},i_{0}}e^{-\beta W} P(i_{\tau},i_{1},i_{0})\nn\\
\end{equation}
using Eq.(\ref{work}), Eq.(\ref{jprob2}) and Eq.(\ref{iprob})
\begin{align}
 \langle e^{-\beta W}\rangle 
&=\sum_{i_{0},i_{1},i_{\tau}}e^{-\beta(E_{i_{\tau}}(\lambda(\tau))-E_{i_{0}}(\lambda(0))}  |\langle i_{1}|U_{\lambda}(t_{1},0)|i_{0}\rangle|^{2}|\langle i_{\tau}|U_{\lambda}(\tau,t_{1})|i_{1}\rangle|^{2} \dfrac{e^{-\beta E_{i_{0}}(\lambda (0)) }}{Z_{0}}\nn\\
&=\sum_{i_{0},i_{1},i_{\tau}}\dfrac{e^{-\beta E_{i_{\tau}}(\lambda (\tau)) }}{Z_{0}}\langle i_{1}|U_{\lambda}(t_{1},0)|i_{0}\rangle\langle i_{0}|U^{\dagger}_{\lambda}(t_{1},0)|i_{1}\rangle|\langle i_{\tau}|U_{\lambda}(\tau,t_{1})|i_{1}\rangle|^{2}\nn\\ 
&=\sum_{i_{1},i_{\tau}}\dfrac{e^{-\beta E_{i_{\tau}}(\lambda (\tau)) }}{Z_{0}} 
\langle i_{\tau}|U_{\lambda}(\tau,t_{1})|i_{1}\rangle\langle i_{1}|U^{\dagger}_{\lambda}(\tau,t_{1})|i_{\tau}\rangle\nn\\
&=\sum_{i_{\tau}}\dfrac{e^{-\beta E_{i_{\tau}}(\lambda (\tau)) }}{Z_{0}}
=\dfrac{Z_{\tau}}{Z_{0}}
=e^{-\beta \Delta F}.
\end{align}
In the above simplification we have used completeness relation, normalization condition and unitarity of $U_{\lambda}$ as in section \ref{sec:JE}. Thus, we find that the JE remains unaffected even if measurements are performed on the system in-between $(0,\tau)$. The above treatment can be readily generalized to the case of multiple measurements (see appendix A). Even though the form of JE is not altered in the presence of measurements, the statistics of the work performed on the system changes (strongly influenced by measurements). This is due to the fact that path probabilities for a given value of work are modified in presence of measurements. This is clearly illustrated in \cite{han11a}, wherein work distribution has been calculated for the Landau-Zener model in presence of measurement.
\section{Extended JE in presence of feedback}\label{sec:JEf}
The extended JE in presence of feedback has been given by Sagawa and Ueda for both the classical \cite{sag10,sag11} and  the quantum \cite{tas11} cases.
Feedback means that system will be controlled by the the measurement output. After each measurement, the protocol is changed accordingly. Suppose initial protocol was $\lambda(t)$; at time $t_{1}$ a measurement of some observable $A$ is performed on the system and outcome $y_1$ is obtained. We then modify our protocol from $\lambda_0(t)$ to $\lambda_{y_1}(t)$ and evolve the system up to time $\tau$. We assume that  the intermediate measurements can involve errors that are purely classical in nature. The error probability is given by $p(y_1|i_1)$, where $|i_1\ra$ is the system's actual state. The final value of the protocol $\lambda_{y_1}(\tau)$ depends on $y_1$ and hence equilibrium free energy at the end of the protocol depends on $y_1$. The mutual information between actual state $|i_{1}\rangle$ and measured value $y_{1}$ is 
\begin{equation}
 I=\ln\dfrac{p(y_{1}|i_{1})}{p(y_{1})}
\label{inform}
\end{equation}
The mutual information $I$ quantifies a change in uncertainty about the state of the system upon making measurement \cite{hor10}. Note that $I$ can be positive or negative for a given realization; however, $\la I \ra$ is always positive.
The probability of the  state trajectory $|i_{0}\rangle\rightarrow|i_{1}\rangle\rightarrow|i_{\tau}\rangle$ with single measurement is
\begin{align}
 P(i_{\tau},i_{1},i_{0},y_{1})
&=p(i_{\tau}|i_{1}) p(y_{1}|i_{1}) p(i_{1}|i_{0})p(i_{0})\nn\\
&=|\langle i_{\tau}|U_{\lambda_{y_{1}}}(\tau,t_{1})|i_{1}\rangle|^{2} p(y_{1}|i_{1}) 
|\langle i_{1}|U_{\lambda}(t_{1},0)|i_{0}\rangle|^{2}p(i_{0}).
\label{jprob3}
\end{align}
Now we have,
\begin{align}
 &\langle e^{-\beta (W-\Delta F)-I}\rangle
=\int dy_{1}\sum _{i_{\tau},i_{1},i_{0}} P(i_{\tau},i_{1},i_{0},y_{1})e^{-\beta (W-\Delta F(y_1))-I}
\end{align}
 Substituting the expressions of joint probability $P(i_{\tau},i_{1},i_{0},y_{1})$ (Eq.(\ref{jprob3})), work W (Eq.(\ref{work})), Free energy difference $\Delta F=\frac{Z_0}{Z_{\tau}(y_1)}$, and mutual information $I$ (Eq.(\ref{inform})) and simplifying we get
\begin{align}
\langle e^{-\beta (W-\Delta F)-I}\rangle
&=\int dy_{1}\sum _{i_{\tau},i_{1},i_{0}}
|\langle i_{\tau}|U_{\lambda_{y_{1}}}(\tau,t_{1})|i_{1}\rangle|^{2} 
|\langle i_{1}|U_{\lambda}(t_{1},0)|i_{0}\rangle|^{2}p(y_{1}) 
   \dfrac{e^{-\beta E_{i_{\tau}}(\lambda _{y_{1}}(\tau)) }}{Z_{\tau}(y)}\nn\\
&=\int dy_{1}\sum _{i_{\tau},i_{1},i_{0}}
|\langle i_{\tau}|U_{\lambda_{y_{1}}}(\tau,t_{1})|i_{1}\rangle|^{2} 
\langle i_{1}|U_{\lambda}(t_{1},0)|i_{0}\rangle \langle i_{0}|U^{\dagger}_{\lambda}(t_{1},0)|i_{1}\rangle
p(y_{1}) \dfrac{e^{-\beta E_{i_{\tau}}(\lambda _{y_{1}}(\tau)) }}{Z_{\tau}(y)}\nn\\
&=\int dy_{1}\sum _{i_{\tau},i_{1}}
|\langle i_{\tau}|U_{\lambda_{y_{1}}}(\tau,t_{1})|i_{1}\rangle|^{2} 
p(y_{1}) \dfrac{e^{-\beta E_{i_{\tau}}(\lambda _{y_{1}}(\tau)) }}{Z_{\tau}(y)}\nn\\
&=\int dy_{1}\sum _{i_{\tau},i_{1}}
\langle i_{\tau}|U_{\lambda_{y_{1}}}(\tau,t_{1})|i_{1}\rangle
\langle i_{1}|U^{\dagger}_{\lambda_{y_{1}}}(\tau,t_{1})|i_{\tau}\rangle
p(y_{1}) \dfrac{e^{-\beta E_{i_{\tau}}(\lambda _{y_{1}}(\tau)) }}{Z_{\tau}(y)}\nn\\
&=\int dy_{1}p(y_{1})\sum _{i_{\tau}}\dfrac{e^{-\beta E_{i_{\tau}}(\lambda _{y_{1}}(\tau)) }}{Z_{\tau}(y)}\nn\\
&=\int dy_{1}p(y_{1})=1.
\label{gJE}
\end{align}
In second and fourth step, the modulus squared terms have been rewritten in expanded form and completeness relation is used.
The above relation (\ref{gJE}) constitutes the extended JE in the presence of information.
Using Jensen's inequality, one arrives at the generalized version of the second law in presence of feedback:
\begin{equation}
\la W\ra \ge \la\Delta F\ra - k_BT\la I\ra,
\end{equation}
where the average mutual entropy $\la I\ra$ is always non-negative on account of being a relative entropy \cite{cov}. Thus, the lower bound of the mean work done on the system can be lowered by a term that is proportional to the average of the mutual information. In other words, with the help of an efficiently designed feedback, we can extract more work from the system. The above treatment can be readily extended to the case of multiple measurements between (0,$\tau$)  not necessarily at equal intervals of time. This is given in appendix B.  
\section{Generalized JE and efficacy parameter in presence of feedback}\label{sec:effpara}
The efficacy parameter $\gamma$ \cite{sag10,sag11,sag08} provides a measure of how efficiently our feedback is able to extract work from the system. It is defined as
\begin{align}
 \gamma \equiv\langle e^{-\beta (W-\Delta F)}\rangle 
=\int dy_{1}\sum _{i_{\tau},i_{1},i_{0}} P(i_{\tau},i_{1},i_{0},y_{1})e^{-\beta (W-\Delta F)}.
\end{align}
Here we have assumed single intermediate measurement. Substituting the expressions of joint probability $P(i_{\tau},i_{1},i_{0},y_{1})$ (Eq.(\ref{jprob3})), work W (Eq.(\ref{work})), Free energy difference $\Delta F=\frac{Z_0}{Z_{\tau}(y_1)}$, and information $I$ (Eq.(\ref{inform})), we get
\begin{align}
\langle e^{-\beta (W-\Delta F)}\rangle 
&=\int dy_{1}\sum _{i_{\tau},i_{1},i_{0}}
|\langle i_{\tau}|U_{\lambda_{y_{1}}}(\tau,t_{1})|i_{1}\rangle|^{2} 
|\langle i_{1}|U_{\lambda}(t_{1},0)|i_{0}\rangle|^{2}p(y_{1}|i_{1}) 
   \dfrac{e^{-\beta E_{i_{\tau}}(\lambda _{y_{1}}(\tau)) }}{Z_{\tau}(y)}\nn\\
&=\int dy_{1}\sum _{i_{\tau},i_{1},i_{0}}
|\langle i_{\tau}|U_{\lambda_{y_{1}}}(\tau,t_{1})|i_{1}\rangle|^{2} 
\langle i_{1}|U_{\lambda}(t_{1},0)|i_{0}\rangle \langle i_{0}|U^{\dagger}_{\lambda}(t_{1},0)|i_{1}\rangle
p(y_{1}|i_{1}) \dfrac{e^{-\beta E_{i_{\tau}}(\lambda _{y_{1}}(\tau)) }}{Z_{\tau}(y)}\nn\\
&=\int dy_{1}\sum _{i_{\tau},i_{1}}
|\langle i_{\tau}|U_{\lambda_{y_{1}}}(\tau,t_{1})|i_{1}\rangle|^{2} 
p(y_{1}|i_{1}) \dfrac{e^{-\beta E_{i_{\tau}}(\lambda _{y_{1}}(\tau)) }}{Z_{\tau}(y)}.
\end{align}
For further calculations we need to take into account time reversed path. For this we introduce time reversal operator $\Theta$ with the properties $\Theta^{\dagger}=\Theta$ and
$\Theta^{\dagger} \Theta=1$. Let $|i_{0}^{*}\rangle$ denote the time reversed state of $|i_{0}\rangle$, i.e, $|i_{0}^{*}\rangle=\Theta|i_{0}\rangle$. It follows \cite{lah11}
\begin{equation}
 \Theta U_{\lambda_{y_{1}}}(\tau,t_{1})\Theta^{\dagger}=U_{\lambda^{\dagger}_{y_{1}}}(\tilde{\tau},\tilde{t}_{1})
\end{equation}
where $\tilde{t}=\tau -t$, i.e, the time calculated along reverse process.
We assume time-reversibility of measurements,  $p(y^{*}_{1}|i^{*}_{1})=p(y_{1}|i_{1})$ \cite{sag08}. As $i^*$ and $i$ have one to one correspondence, the summation over  $i_1$, $i_{\tau}$ is equivalent to that over $i_1^*$, $i_{\tau}^*$. We get
\begin{align}
\langle e^{-\beta (W-\Delta F)}\rangle 
&=\int dy_{1}\sum _{i^*_{\tau},i^*_{1}}
|\langle i_{\tau}|\Theta^{\dagger} \Theta U_{\lambda_{y_{1}}}(\tau,t_{1})\Theta^{\dagger} \Theta|i_{1}\rangle|^{2} 
p(y_{1}|i_{1}) \dfrac{e^{-\beta E_{i_{\tau}}(\lambda _{y_{1}}(\tau)) }}{Z_{\tau}(y)},\nn\\
&=\int dy_{1}\sum _{i^*_{\tau},i^*_{1}}|\langle i^{*}_{\tau}|U_{\lambda^{\dagger}_{y_{1}}}(\tilde{\tau},\tilde{t}_{1})|i^{*}_{1}\rangle|^{2}
p(y^{*}_{1}|i^{*}_{1}) \dfrac{e^{-\beta E_{i_{\tau}}(\lambda _{y_{1}}(\tau)) }}{Z_{\tau}(y)}\nn\\
&=\int dy_{1}\sum _{i^*_{\tau},i^*_{1}}|\langle i^{*}_{1}|U^{\dagger}_{\lambda^{\dagger}_{y_{1}}}(\tilde{\tau},\tilde{t}_{1})|i^{*}_{\tau}\rangle|^{2}
p(y^{*}_{1}|i^{*}_{1}) \dfrac{e^{-\beta E_{i_{\tau}}(\lambda _{y_{1}}(\tau)) }}{Z_{\tau}(y)}\nn\\
&=\int dy_{1}\sum _{i^*_{\tau},i^*_{1}}|\langle i^{*}_{1}|U_{\lambda^{\dagger}_{y_{1}}}(\tilde{t}_{1},\tilde{\tau})|i^{*}_{\tau}\rangle|^{2}
p(y^{*}_{1}|i^{*}_{1}) \dfrac{e^{-\beta E_{i_{\tau}}(\lambda _{y_{1}}(\tau)) }}{Z_{\tau}(y)}\nn\\
&=\int dy_{1}\sum _{i^*_{\tau},i^*_{1}} P_{\lambda^{\dagger}_{y_{1}}}(i^{*}_{1}|i^{*}_{\tau})p(y^{*}_{1}|i^{*}_{1})P(i_{\tau}).
\label{kuchi}
\end{align}
where 
\begin{equation}
 P_{\lambda^{\dagger}_{y_{1}}}(i^{*}_{1}|i^{*}_{\tau})=|\langle i^{*}_{1}|U_{\lambda^{\dagger}_{y_{1}}}(\tilde{t}_{1},\tilde{\tau})|i^{*}_{\tau}\rangle|^{2},
\end{equation}
is the conditional probability of time reversed trajectory from state $|i^{*}_{\tau}\rangle$ to $|i^{*}_{1}\rangle$. We also have
\begin{equation}
 P(i^{*}_{\tau})= P(i_{\tau})=\dfrac{e^{-\beta E_{i_{\tau}}(\lambda _{y_{1}}(\tau)) }}{Z_{\tau}(y_1)},
\label{revprob}
\end{equation}
which is the initial probability distribution of the time reversed process with fixed protocol $\lambda^{\dagger}_{y_{1}}(\tau)$. Substituting Eq.(\ref{revprob}) in Eq.(\ref{kuchi}) and using Bayes' theorem we get
\begin{align}
\gamma=\langle e^{-\beta (W-\Delta F)}\rangle 
=\int dy_{1}\sum _{i^*_{1}}p(y^{*}_{1}|i^{*}_{1})P_{\lambda^{\dagger}_{y_{1}}}(i^{*}_{1})
=\int dy_{1}P_{\lambda^{\dagger}_{y_{1}}}(y_1^{*}).
\end{align}
The physical meaning of the efficacy parameter is apparent now: it is the total probability of observing time-reversed outcomes along time-reversed protocols. Thus expression for the efficacy parameter remains the same as in the classical case. For multiple measurements, efficacy parameter is given by $\gamma=\int dy_{1}\cdots dy_n P_{\lambda^{\dagger}}(y^{*}_1\cdots y_n^*)$. The derivation is simple and we are not reproducing it here.

 In conclusion we have shown that the quantum extension of JE with multiple measurements and measurement accompanied feedback and quantum efficacy parameter retain same expressions as in the classical case. This is mainly due to performed measurements being of von Neumann projective type accompanied by classical errors, and system being isolated.  We have also shown that in quantum case, entropy production fluctuation theorems retain the same form as in the classical case with measurement and feedback. The results will be published elsewhere.

\section{Acknowledgement}

One of us (AMJ) thanks DST, India for financial support.
\appendix

\section{JE in presence of multiple measurements} \label{JEmm}
We ~consider~ n number of intermediate ~measurements ~of any observable ~being performed at time $t_{1},t_{2},....,t_{n}$ and the system colapses to it's corresponding eigenstate at
 $|i_{1}\ra,|i_{2}\ra,...|i_{n}\ra$ respectively. Here we have considered the system evolves with the predetermined protocol $\lambda(t)$. The probability of the corresponding  state trajectory 
\begin{align}
 P(i_{\tau},....i_{2},i_{1},i_{0})
&=p(i_{\tau}|i_{n})...p(i_{2}|i_{1})p(i_{1}|i_{0})p(i_{0})\nn\\
&=|\langle i_{\tau}|U_{\lambda}(\tau,t_{n})|i_{n}\rangle|^{2}
...|\langle i_{2}|U_{\lambda}(t_{2},t_{1})|i_{1}\rangle|^{2}|\langle i_{1}|U_{\lambda}(t_{1},0)|i_{0}\rangle|^{2}p(i_{0})
\end{align}
\begin{align}
 &\langle e^{-\beta W}\rangle \nn\\
&=\sum_{i_{0},i_{1},...,i_{\tau}} e^{-\beta(E_{i_{\tau}}(\lambda(\tau))-E_{i_{0}}(\lambda(0)))} P(i_{\tau},....i_{2},i_{1},i_{0})\nn\\
&=\sum_{i_{0},i_{1},...,i_{\tau}}e^{-\beta(E_{i_{\tau}}(\lambda(\tau))-E_{i_{0}}(\lambda(0))}  |\langle i_{\tau}|U_{\lambda}(\tau,t_{n})|i_{n}\rangle|^{2}
...|\langle i_{2}|U_{\lambda}(t_{2},t_{1})|i_{1}\rangle|^{2}|\langle i_{1}|U_{\lambda}(t_{1},0)|i_{0}\rangle|^{2}p(i_{0}).
\end{align}
Using completeness and normalization  of eigenstates $|i_{0}\ra,|i_{1}\ra,...|i_{n}\ra$ and unitarity of evolution, we get after  simplification
\begin{align}
\langle e^{-\beta W}\rangle 
=\sum_{i_{\tau}}\dfrac{e^{-\beta E_{i_{\tau}}(\lambda (\tau)) }}{Z_{0}} 
=\dfrac{Z_{\tau}}{Z_{0}}=e^{-\beta \Delta F}.
\end{align}
Thus JE retain the same classical form even in presence of multiple measurements.
\section{multiple measurement and feedback}
Let ~the ~outcome ~of measurement ~values at ~time $t_{1},t_{2},...,t_{n}$ is $y_{1},y_{2},...,y_{n}$  with ~a classical measurement error $p(y_{1}|i_{1}),p(y_{2}|i_{2})$ $,...,p(y_{n}|i_{n})$ ~ respectively ~ when ~actual ~intermediate~ states~ are~ $|i_1\ra $, $|i_2\ra \cdots|i_n\ra $.~ The state $|i_0\ra $ and $|i_{\tau}\ra $ are observed projected eigenstates of energy observable in the beginning and end of the protocol. The total path probability can be expressed as
\begin{align}
 P(i_{\tau},..,i_{1},i_{0},y_{n},..,y_{1})=|\langle i_{\tau}|U_{\lambda_{y_{n}}}(\tau,t_{1})|i_{n}\rangle|^{2}...p(y_{2}|i_{2})|\langle i_{2}|U_{\lambda_{1}}(t_{2},t_{1})|i_{1}\rangle|^{2}p(y_{1}|i_{1}) \nn\\
\hspace{3cm}\times|\langle i_{1}|U_{\lambda}(t_{1},0)|i_{0}\rangle|^{2}p(i_{0}).
\end{align}
Now, 
\begin{align}
&\langle e^{-\beta (W-\Delta F)-I}\rangle
=\int dy_n,..,dy_{1}\sum _{i_{\tau},..,i_{1},i_{0}}P(i_{\tau},..,i_{1},i_{0},y_{n},..,y_{1}) e^{-\beta (W-\Delta F)-I}.
\end{align}
Substituting value of work $W$ (Eq.(\ref{work})), mutual information $I=\ln \frac{p(y_n|i_n)...p(y_2|i_2)p(y_1|i_1)}{p(y_n,...y_2,y_1)}$, Free energy difference $\Delta F=\frac{Z_0}{Z_{\tau}(y_n)}$,  and simplifying we get 
\begin{align}
 &\langle e^{-\beta (W-\Delta F)-I}\rangle\nn\\
&=\int dy_n\cdots dy_{1}\sum _{i_{\tau},..,i_{1},i_{0}}|\langle i_{\tau}|U_{\lambda_{y_{n}}}(\tau,t_{1})|i_{n}\rangle|^{2}...|\langle i_{2}|U_{\lambda_{1}}(t_{2},t_{1})|i_{1}\rangle|^{2}
|\langle i_{1}|U_{\lambda}(t_{1},0)|i_{0}\rangle|^{2}\nn\\
&\hspace{8cm}\times p(y_n,...,y_2,y_1) \dfrac{e^{-\beta E_{i_{\tau}}(\lambda _{y_{n}}(\tau)) }}{Z_{\tau}(y_n)},\nn\\
&=\int dy_n\cdots dy_{1}p(y_n,...,y_2,y_1)\sum _{i_{\tau}} \dfrac{e^{-\beta E_{i_{\tau}}(\lambda _{y_{n}}(\tau)) }}{Z_{\tau}(y_n)},\nn\\
&=\int dy_n\cdots dy_{1}p(y_n...y_2,y_1) =1.
\end{align}
This is the extended quantum JE in the presence of multiple measurements accompanied by feedback.

\end{document}